# Chapter 2 Introduction to Big data Technology

Bilal Abu-Salih[1], Pornpit Wongthongtham[2]
Dengya Zhu[3] , Kit Yan Chan[3] , Amit Rudra[3]

[1] The University of Jordan
[2] The University of Western Australia
[3] Curtin University

**Abstract**: Big data is no more "all just hype" but widely applied in nearly all aspects of our business, governments, and organizations with the technology stack of AI. Its influences are far beyond a simple technique **innovation** but involves all rears in the world. This chapter will first have historical review of big data; followed by discussion of characteristics of big data, i.e. from the 3V's to up 10V's of big data. The chapter then introduces technology stacks for an organization to build a big data application, from infrastructure/platform/ecosystem to constructional units and components. Finally, we provide some big data online resources for reference.

### 2.1 Introduction

The ability to exploit the ever-growing amounts of business-related data will allow to comprehend what is emerging in the world. In this context, Big Data is one of the current major buzzwords [1]. Big Data (BD) is the technical term used in reference to the vast quantity of heterogeneous datasets which are created and spread rapidly, and for which the conventional techniques used to process, analyse, retrieve, store and visualise such massive sets of data are now unsuitable and inadequate. This can be seen in many areas such as sensor-generated data, social media, uploading and downloading of digital media.

Advanced, unconventional and adaptable analytics are required to address the challenges of managing and analysing a wide variety of BD islands [2] which are expanding exponentially as a result of the huge amount of data being generated by tracking sensors, social media, transaction records, and metadata to name just a few of the many sources of data. For example, every one second there occur 8,910 new Tweets, 89,845GB usage of internet traffic, 81,734 Google searches, etc. [3] [3] [24][23]. Furthermore, as depicted in Figure 2.1, the digital universe data is expected to grow by a factor of 10 by 2020s [4].

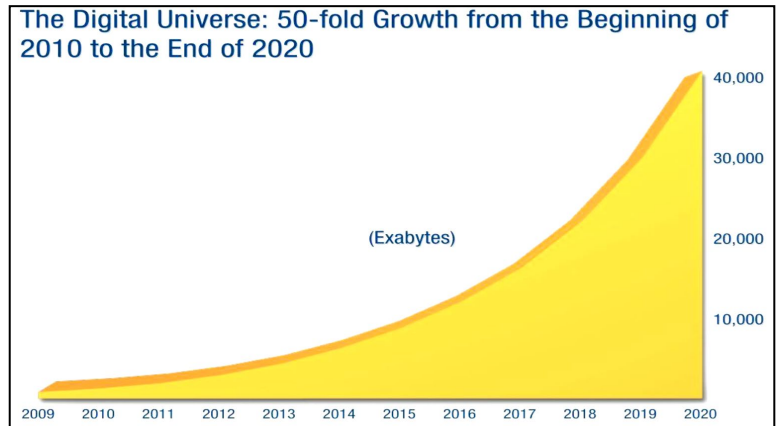

**Figure 2.1: Digital Growth[1]**

These massive amounts of generated data have implications for businesses and how they operate [5, 6]. In particular, business firms have over-indulged in providing information which stimulates the decision makers to adapt to the rapid increase in data volume. However, amongst the 85 percent of those that have aspired to become data-driven companies, only 37 percent have proven to be successful [7]. The information overload[2] (a.k.a. infoxication [8] or infobesity [9]) has made the process of decision making much more difficult. This has led to the need for a close scrutiny of internal business processes, and a review of the tools used to collect, transfer, store and analyse the flood of data generated by a company's internal and external data sources.

**More Unstructured Data and Less Structured Data.**

Data is everywhere, presented in various formats, and is collected from many heterogeneous resources. Business Intelligence (BI) applications are more focused on structured data and support decision-makers by providing meaningful information from extracted data mainly coming from day-to-day operational information systems and structured external data sources [10]. With the rapid increase in the amount of unstructured data, traditional data warehouses cannot be the sole source of data analytics. If 20 percent of data available in an organisation is mainly structured data [11], the unstructured data accounts for 80 percent of the total data that the organisation encounters. Examples of unstructured data include free text, emails, images, audio files, streaming videos, and many other data types [12].

The size of large data can range from several terabytes to petabytes and even exabytes [13]. Thus, there is an inevitable need to develop platforms that deal with this large scale of data. This has driven many companies to examine their current information and communication technology infrastructure, to strengthen the expertise of their employees, and to prepare them to benefit from the systems of sophisticated unstructured data analysis correctly and effectively [14]. In fact, The continuous investment in the world of Big data is broadcasted to several firms with dissimilar sizes, and operating in manifold business sectors; 75 percent of the 400 companies surveyed by Gartner reported that their BD investment had been either started or anticipated to start within the upcoming years [15]. In this context, organisations have ever become much decisive to cope with the enormous influence of BD in many aspects of business practices [16]. This highlights the need to review the tools used to collect, transfer, store, and analyse such massive amounts of data [20-17].

---

[1] Source: IDC's Digital Universe Study, sponsored by EMC, December 2012

[2] The term "Information overload" has been popularized by Alvin Toffler in his book, Future Shock, 1971.

Another survey carried out by Gartner [21] foresee that by 2020 business firms will continue their spacious investment on social and big data analytics, and they will provide platforms for users to access curated and credible data collected from internal and external data sources. In addition, Deloitte survey [22] emphasizes the importance of data analytics in the business domain; around 75 percent of all respondents believe that the adoption of the continuous propagation of data will benefit their business strategies; and 96 percent of the respondents consider data analytics as an added value for their businesses in the coming three years.

### 2.2 History of Big Data

The notion of Big data comes before the advances in databases technologies and from the need for solutions to handle the huge deluge of datasets and, therefore, the lack of sufficient storage capacity [23]. As depicted in Figure 2.2 the notion of Big data has evolved through the past decades where each decade is described in terms of computer disc space, from Megabyte (MB) in 1970s to Exabyte (EB) which was introduced in 2011.

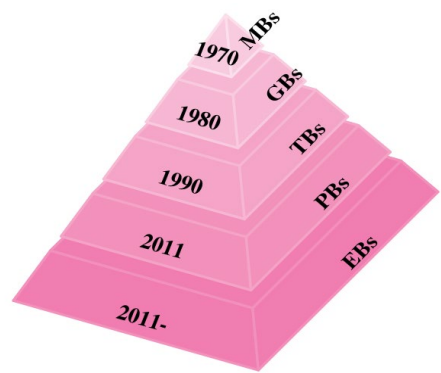

**Figure 2.2: Brief history of big data [23, 24]**

Despite numerous attempts to provide a consensus definition to the Big data, the term confronts an overall uncertainty among researchers in dissimilar disciplines, and it has been incorporated as a new and sophisticated apparatus for research [25]. Although the term has been extensively used in the recent decade, the term itself in its true essence is not quite new and it is argued that the term was coined back in 1990s [26]. Nevertheless, the term has been deliberated since 2000 and was associated with statistics domain, and thus it was used to describe "the explosion in the quantity (and sometimes, quality) of available and potentially relevant data, largely the result of recent and unprecedented advancements in data recording and storage technology" [27]. In the following years, the usage of the term has notably extended to many domains. One of the most comprehensive and well recognised definition is the one provided by the European Commission:

> *"Large amounts of different types of data produced from various types of sources, such as people, machines or sensors. This data includes climate information, satellite imagery, digital pictures and videos, transition records or GPS signals. Big Data may involve personal data: that is, any information relating to an individual, and can be anything from a name, a photo, an email address, bank details, posts on social networking websites, medical information, or a computer IP address"* [28]

The Big Data Framework organization [29] attempts to categorise the development of Big data to three main phases; (i) **Phase 1.0**: Big data was mainly described by the data storage and analytics, and it was an extension to the modern database management systems and data warehousing technologies; (ii) **Phase 2.0**: With the uprising of Web 2.0, and the propagation of semi-structured and unstructured content, the notion of Big data has evolved to embody advanced technical solutions to extract meaningful information from dissimilar and heterogeneous data formats; (iii) **Phase 3.0**: with the emergence of smartphones and mobile devices, sensor data, Internet of Things (IoT), wearable devices, to many more data generators, Big Data has entered a new era and has drawn a new horizon with a new range of opportunities [29]. A summary to the three proposed Big data phases is borrowed and presented in Figure 2.3.

| BIG DATA PHASE 1 | BIG DATA PHASE 2 | BIG DATA PHASE 3 |
|---|---|---|
| Period: 1970-2000 | Period: 2000-2010 | Period: 2010-present |
| DBMS-based, structured content: <br> • RDBMS & data warehousing <br> • Extract Transfer Load <br> • Online Analytical Processing <br> • Dashboards & scorecards <br> • Data mining & statistical analysis | Web-based, unstructured content <br> • Information retrieval and extraction <br> • Opinion mining <br> • Question answering <br> • Web analytics and web intelligence <br> • Social media analytics <br> • Social network analysis <br> • Spatial-temporal analysis | Mobile and sensor-based content <br> • Location-aware analysis <br> • Person-centered analysis <br> • Context-relevant analysis <br> • Mobile visualization <br> • Human-Computer-Interaction |

**Figure 2.3: Three phases of Big data [29]**

Big data is anecdotally seen as the conjunction of two spheres; data storage and data analysis. This perspective poses a question on the novelty of the term considering that Big data in this essence is not different from the conventional data analytics techniques. In fact as indicated by Ward and Barker [30], although the term "Big" designates complexity, sophistication and challenge, the term has been predominantly used to invite quantities which has diverged the term from its factual meaning, thereby establishing such lack of consensus on the definition and kept shrouded by much conceptual vagueness [31]. This indeed requires shedding the light on the key features that distinguish Big data from the traditional and conventional data analytics [32-36]. Next section provides a brief depiction to Big data features.

### 2.3 Characteristics of Big Data

The diverse depictions to the Big data problem have naturally led to provide a plethora of technical perceptions on the Big data paradigm. This has opened doors to several descriptions to Big data in dissimilar contexts toward better understanding to Big data paradigm, its challenges and advantages, and hence, obtaining the true value of it [37-43]. To differentiate it from the traditional data processing systems, Big data has been amply characterised by the well-known 3Vs (*Volume, Velocity* and *Variety*) [44]. However, the businesses have found these three dimensions less adequate to tackle Big data properly and thus insufficient to provide hoped-for valuable insights. Hence, with seven more features have been proposed, a consolidated depiction to the Big data problem has been obtained. These features are mainly; *V*eracity, *Variability, Value, Validity, Vulnerability, Volatility and Visualization.* The following is a brief discussion on the 10Vs of Big data.

*Volume:* refers to the vast increase in the data growth. This is evident as more than 90% of the data we encouter was produced recently [45]. In fact, more than 2.5 quinilion($10^{18}$) bytes are created daily since even as earlier as 2013 from every post, share, search, click, stream, and many more data producers [46]. Although this abundance of data poses a challnge on storage capacity, this challenge is less stimulating due to the advanced storage technologies as well as the decrease in the cost of computer storage acquisition [38]. However, analysis of such vast data islands is the actual challenge considering the heterogenity nature of data.

*Velocity:* represents the accumulation of data in high speed, near real-time and real-time from dissimilar data sources. Velocity embodies also the temporal and latency qualitites. The velocity of data requires advanced solutions that are able to store, process, manage and analysis streams of heteroginous data and infer value on motion [47]. Figure 2.4 shows a depiction on how much data is generated in a minute from various resources in 2019.

*Variety:* involves collecting data from various resources and in fuzzy and heterogeneous types [48]. This icludes importing data in dissimlar formats, namely structured (tables reside in relational databases – RDBMS, etc.), semi-structured (email, XML, JSON, and other markup languages, etc.) and unstructured (text, pictures, audio files, video, sensor data, etc.). Variety indicates also importing datasets from differet repositories. For example, business firms conduct their analysis on data collected from day-to-day structural databases, and also from social unstrctured data obtained from social media.

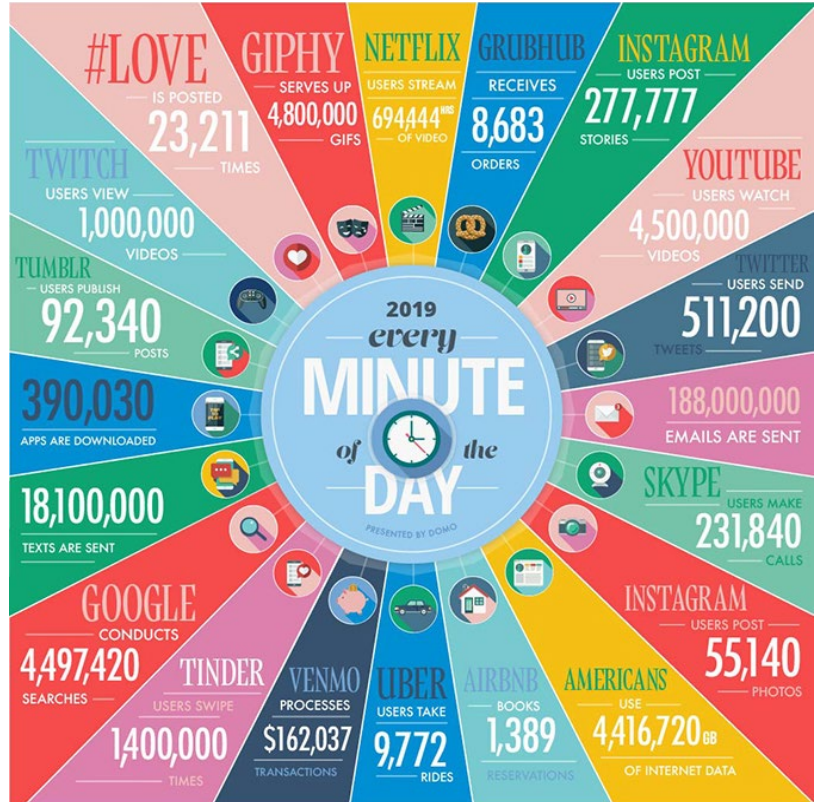

**Figure 2.4: How much data is generated in each minute [3]**

**Veracity:** refers to the provenance, accuracy, and correctness of data. It also refers to objectivity vs subjectivity, truthfulness vs deception and credibility vs implausibility [49]. Demchenko et al. [50] presented multiple factors to ensure the *veracity* of Big data. These factors include but are not limited to: (i) trustworthiness of data origin; (ii) reliability and security of data store; and (iii) data availability. The list established by Demchenko et al. could be enhanced by including a further two essential aspects: *correctness* and *consistency*. Although data origin and store are critical, the trustworthiness of the source does not guarantee data correctness and consistency. Data cleansing and integration should be incorporated to ensure the *veracity* of data as well. For example, in the context of SBD (Social Big Data), the data collected from social media should be examined to infer high-quality content and eliminate the poor-quality data for further data analysis. Poor data quality has a major negative impact on the data analysis process, and the output will lack credibility and trustworthiness. Hence, organisations should understand how to extract valuable and veracious data.

---

[3] Source: https://www.domo.com/learn/data-never-sleeps-7

*Variability:* refers to variance in meaning, number of inconsistences, multitude of data dimensions, and inconsistent data receiving speeds [51-54]. Several attempts have been proposed to address this problem; the incorporation of semantic analysis in SBD, for example, reduces the ambiguity of SBD by clarifying the actual context of the users' content. This mitigates the *variability* of Big data [54] [55].

*Validity*: refers to the "data are shown (or known) to be an accurate indicator of the claim being made" [56]. Validity differs from the veracity in that the validity does "mean the correctness and accuracy of data with regard to the intended usage" [57]. In other word, data can be trustworthy, thus satisfy the veracity aspect. Yet, poor interpretation to the data might lead to unintended use. Moreover, the same *veracious* data can be *valid* to be used in one application and *invalid* for a different one [57].

*Vulnerability*: refers to the security of the collected datasets that will be used for later analysis [58]. It also denotes the flaws in the system which permits malicious activities to be conducted on the collected datasets. Hence, the acquisition of datasets should ensure capacity to provide immune systems able to protect the collected data from breaches [59]. For example, around 165 million user accounts of LinkedIn were leaked in 2012, and data of 500 million customers of Marriott International were stolen in 2018 [60].

*Volatility*: refers to time up which data is valid to be stored/used before it becomes obsolete or no longer relevant [61]. It is crucial dimension since cost of storage and maintenance augments with longer Big data retention [57].

*Visualisation:* refers to the ability to present Big data into a visual context, such as diagrams, graphs, maps, etc. toward better understanding and interpreting of data [62]. It also assists people and organisations to discover patterns, correlations, trends, relationships and dependencies. Big data visualisation is a powerful tool for decision makers to access, evaluate and interpret massive data in even real time and act upon it [63].

*Value*: represents the outcome product of Big data analysis (i.e. new insights) [50]. The impact of Big data abundance extends beyond business-related data to cope data generated from industrial firms, educational institutions, political and governmental data, healthcare data, and many other industries. The key challenge of Big data analysis is the mining of enormous amounts of data in the quest for hoped for added value. SBD is an important Big data island; thus social data analytics aims to make sense of data and to obtain value from social data.

**Figure 2.5** shows a depiction to the 10Vs of Big data along with highlights for each feature.

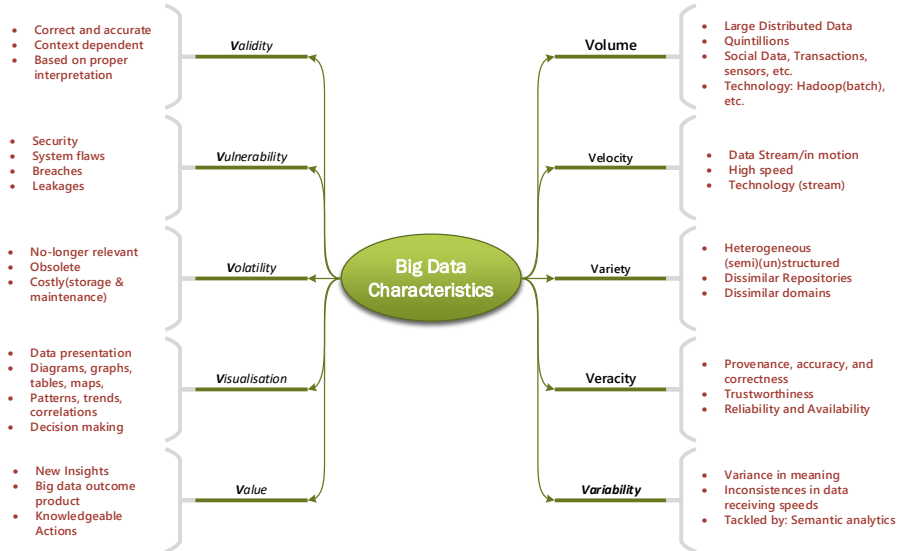

**Figure 2.5: Big data V-features**

## 2.4 Cloud Computing

After discussed the features of big data, the 3V's, 4V's and even 10V's, the next problem is how can we store and process the data and then extract values form the big data. In addition to the traditional relational database systems for structured data storage and processing, different techniques have been developed to handle big data, including but not limited to cloud computing, data lake and enterprise hybrid data cloud. In the following sections, basic concepts of could computing will be introduced at first; then data lake technologies will be discussed and Snowflake's solution will be taken as an example; Lastly, we will introduce Cloudera Data Platform, which is a kind of enterprise level data cloud, or hybrid data cloud that typically is based on a Hadoop Ecosystem for both big data storage and parallel computation.

### 2.4.1 Introduction to Cloud Computing

In 2010, Armbrust et al. [64] envisioned that cloud computing can potentially transform the whole IT industry for both software development and hardware design and purchase. In their definition, cloud computing refers to two things, one is "the applications delivered as services over the Internet", and second, "the hardware and system software in the data centers that provide those services". One example for hardware provider is Amazon's EC2 instance which is a kind of virtual server, and Google's AppEngine is an example of application platform. Microsoft's Azure provide a service that just between that of Amazon and Google. Amrbrust et al. (2010) further identified ten obstacles, which at the same time were opportunities for clouding computing, as listed in Table 2.1.

Armbrust et al. argue that one more obstacle is that decision making of the usage of clouding will add extra management cost which may unpredictable. For example, IT managers must have knowledges and experiences on which clouding providers should be use and which kind of services are best suitable for an organization/business's requirements, and minimize the cost, especially unprecedent extremely high cost of clouding services. A risk analysis model may help business to alleviate potential financial risks and prevent the loss of data – the priceless asset of an organization.

Table 2.1 Top 10 obstacles and opportunities for  growth of cloud computing [64]

| Obstacle | Opportunity |
| --- | --- |
| 1  Availability/Business  Continuity | Use Multiple Cloud Providers |
| 2 Data Lock-In | Standardize APIs; Compatible Software to enable Surge or Hybrid Cloud Computing |
| 3 Data Confidentiality and Auditability | Deploy Encryption, VLANs, Firewalls |
| 4 Data Transfer Bottlenecks | FedExing Disks; Higher Bandwidth Switches |
| 5 Performance Unpredictability | Improved Virtual Machine (VM) Support; Flash Memory; Gang Schedule VMs |
| 6 Scalable Storage | Invent Scalable Store |
| 7  Bugs  in  Large  Distributed Systems | Invent Debugger that relies on Distributed VMs |
| 8 Scaling Quickly | Invent Auto-Scaler that relies on Machine Learning; Snapshots |
| 9 Reputation Fate Sharing | Offer reputation-guarding services like those for email |
| 10 Software Licensing | Pay-for-use licenses |

As Armbrust et al. excluded private data centre in their cloud computing definition, according to America National Institute of Standards and Technology (NIST), clouding computing "is a model for enabling ubiquitous, convenient, on-demand network access to a shared pool of configurable computing resources (e.g., networks, servers, storage, applications, and services) that can be rapidly provisioned and released with minimal management effort or service provider interaction." [65] The document further defined five essential features, four deployment models and three service models.

The three service models are Software as a Service (SaaS), Platform as a Service (PaaS) and Infrastructure as a Service (IaaS). Deployment models include Public cloud, Private cloud, Community cloud and Hybrid cloud. For a given cloud model, NIST specifies it should have the following characteristics: On-demand self-service, Broad network access, Resource pooling, Rapid elasticity and Measured service.

### 2.4.2 Cloud Computing Service Models

Figure 2.6 demonstrated the level of controls by three different services models, i.e. IaaS, PaaS and SaaS. Note that the controlled layers in this figure are a little different from that defined in [65]. For example, in NIST definition, IaaS consumer has control over operating systems, storage and deployed applications; and they may also control some networking components such as firewalls. According to NIST, the control level of components is relative flexible rather than have a clear boundary.

### Infrastructure as a Service (IaaS)

IaaS uses virtual machines or infrastructure on demand. The service suppliers provide network elements, storage hardware, load balancers, and virtual configurable servers according to users' requirements. Users will take control of software installation and system operation, which will require IT professionals in charge of the routinely system run and maintain of the platform. As shown in the figure below, all components include, and above operating systems are under control by consumers. This service model is highly scalable, no physical hardware purchase and security issues, and pay only virtual machines required.

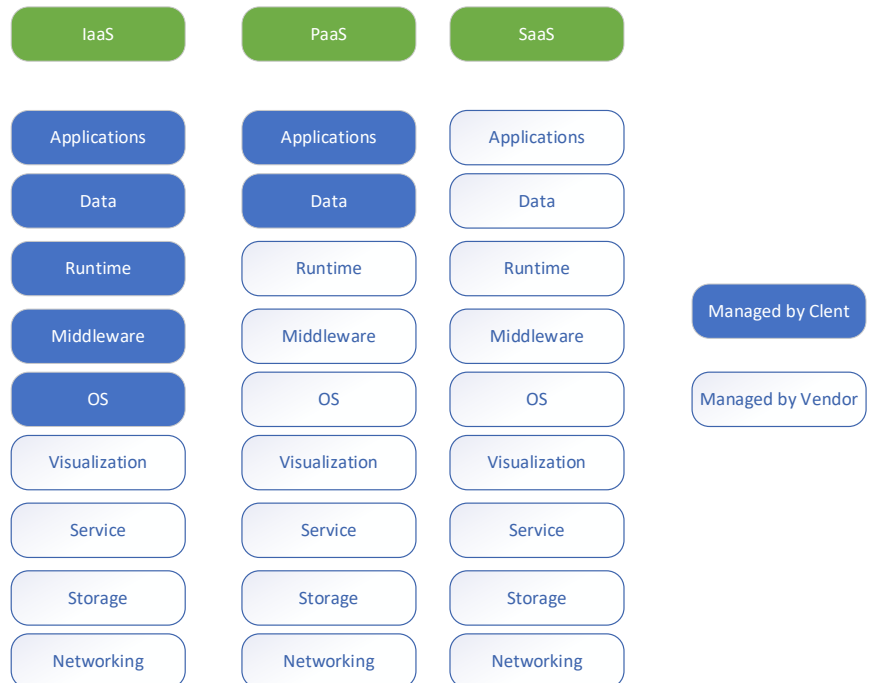

**Figure 2.6 Azure Cloud computer Service models (reproduced from [66])**

### Platform as a Service (PaaS)

As illustrated in Figure 2.6, in PaaS, it is service provider's responsibility to install and maintain operating systems, system software such database management systems. The consumers only focus on their data, applications running on PaaS, and make some necessary configuration settings of operating systems and system software. One obvious advantages of PaaS is that each time when consumers require to increase their hardware configuration, the service provider can upgrade the service promptly, rather than the consumers to reinstall all necessary software by their own, especially when the requirements need to response in a near-real-time fashion, and without interrupting consumers normal operation. Azure Automation and Azure SQL are examples of PaaS.

### Software as a Service (SaaS)

In SaaS, the service providers are responsible for all the controls of the services, and the consumers only use the services provide by SaaS. Consumers have no need to consider from hardware to operation systems, and even how their own data is managed, how and when the application is to be upgraded and improved. What the consumers can do rely on only the configuration of the application provide by the SaaS. This model provides global accessibility and thus easier for collaboration; easier administration and automatic updating/upgrading.

As the increase of available of cloud services, a client may have to access a list of cloud services and thus require identification of each of these services, this may increase the burden of the clients and the cost of an organization to manage these identities. It may also have security concerns if the clients' identities are not properly managed. Identity as a Service (IDaaS) [67] is proposed to handle this issue. IDaaS may include but not limited the following services such as Single sign-on (SSO) services, risk and event monitoring, authentication services, and directory services.

### 2.4.3 Cloud Computing Deployment Models

Deployment models are environments where cloud services are installed. NIST defined four types of deployment models which are introduced as following.

### Public Cloud

Public cloud, as its name indicates, is an infrastructure that provide services open for public users. The owner of a public cloud may be any organization, business, enterprise or combination of them. Current public cloud providers include IT giants such as Amazon and Microsoft which provide cloud services via the Internet.

Public cloud services are usually cost effective, as when a huge number of consumers share the same resources, the overall cost may not too expensive. This public cloud is also more reliable in general because it is run and maintained by IT giants and thus have more human resources and knowledge to handle any technique issues promptly. Users can apply for computing and storage services from public cloud in a flexible manner and don't need to worry about the scalability and location issues.

However, security and privacy are always the concerns of public clouds, and it is not as customable as private cloud.

### Private Cloud

Opposite to public cloud, private cloud infrastructure is provisioned for users within an organization where internal consumers can from different business units. Similar to public cloud, owners and operators of a private cloud may be the organization, a third party, or combination of them, but usually it is owned by the organization, and maintained by third party companies that have the necessary knowledge and skills to keep the private cloud up and run reliably. It may locate on-premise or geographically distributed.

Private cloud provides relatively higher security and privacy, more control, more cost and energy efficiency and more reliability compared with public cloud services. However, it requires to purchase hardware from the cloud setup to satisfy the scalability requirements. It may also require skilled expertise, which may within the organization or via third party company outsourcing.

### Community (cooperative) Cloud

Based on NIST, community cloud infrastructure is suitable for a group of organizations that have shared missions, security requirements, policies, and compliance considerations. This kind of cloud is owned and operated by one or more of the organizations and exists on or off premises.

Community cloud has the same advantages as private cloud, such as more cost effective with cost shared by the community members; more security and privacy protection; and easy for information sharing among the community which is missed by most of the research. For example, in a government cooperative cloud, information within the cloud can be shared more effectively than that if the information is stored in different clouds. This cooperative cloud also reduces the duplications of the same information stored by many government departments; increases the data consistence with a single updated version of information, rather than different departments have their own versions for the same information, which may out-of-date for some reasons, such as not updated in a real-time or near-time manner.

Some concerns on community cloud are that some part of data may be unnecessarily shared among other organizations, issues on allocating responsibilities of governance.

### Hybrid Cloud

This type of infrastructure is composed by two or more afore mentioned cloud infrastructure (public, private, or community clouds), but "are bound together by standardized or proprietary technology that enables data and application portability." [64] In hybrid cloud infrastructure, non-critical data and operations, such as training or learning, are stored and performed on the public cloud; and product data and activities are stored and performed on the private cloud.

This type of cloud may take advantages of other kinds of clouds in terms of scalability, flexibility, cost efficiency and security. But at the same time, it has also the disadvantages of the other type of clouds. In addition, if information needs to be exchanged between private and public clouds, networking may become complex. Attacks on private cloud may come from public cloud, and reliability of private cloud remains a concern because it is not improved by the combining of public cloud.

### 2.2.4 Brief introduction to Amazon AWS, Microsoft Azure, and Google Cloud Platform

Amazon, Microsoft and Google are three major players on the cloud service market. Carey (2020) [68] summarized features provided by the three public cloud service providers; and also listed pros and cons of the cloud services provided by three providers.

**Table 2.2 Cloud service features of Amazon, Microsoft and Google [68]**

| Provider | Pros | Cons |
|---|---|---|
| AWS | Wide range of services<br>Platform configuration options, security, reliability<br>Third party software services | Hybrid cloud support<br>Difficult navigate the large numbers of features<br>Aggressive price |
| Microsoft | Strong footing with organisations<br>Hybrid cloud support<br>More open source oriented<br>Easy transition to Azure from current Microsoft products/services | A recent outage<br>Technical support |
| Google | Open source support innovative | Marketing strategy and policy<br>Immaturity with enterprise accounts<br>No presence in big markets |

### 2.5 Could Data Lakes – Snowflakes

A typical and dominant traditional data processing approach for an organization or enterprise is to build a relational database which contains schemas for tables of structured data, and then extract data from the database for data mining and analytics. To process data of different kinds of types, the following figure (Figure 2.7) illustrates the conventional data lake structure for big data processing.

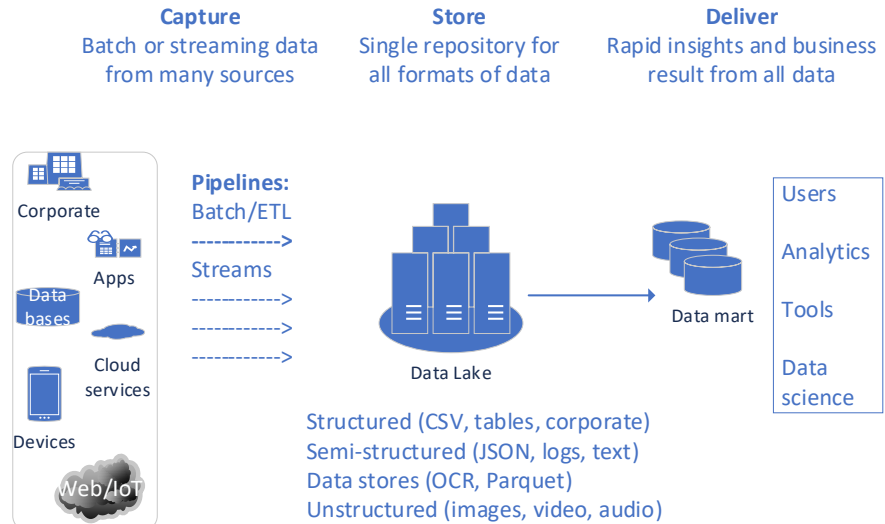

**Figure 2.7 Traditional data lake structure (reproduced from [69])**

The ultimate goal of data lake is to enable users to handle big data in petabytes size without predefined schema. Analogy is that in a huge lake, new water streams via different channels continues, we should able to take samples for analysis. Inspired by the innovative idea, organizations can traditionally build such a data lake based on the Apache Hadoop ecosystem (refer to following section) on-premise.

There are some drawbacks of the traditional data lake processing as summarized by [69]. First the performance of SQL-style query of data stored on HDFS via Apache Hive is relatively slow, although Apache Impala can improve the performance of this kind of queries, it requires more hardware resources and this will have potential effect on the whole performance of the Hadoop Ecosystems. A second drawback is that ETL is a very complex process which involves multiple systems, hard to manage and needs expensive skilled labour to code. Another issue is that data governance is poor, poor resource and access control, poor integration and lack of audit trail. These weaknesses lead the Hadoop-based traditional data lakes became data swamps from there business insights are hard to extract [69]. However, we observed that Hadoop ecosystems are also evolving, and the above-mentioned issues and drawbacks are addressed correspondingly, as we can see in the following session when discussing Cloudera's new hybrid platform.

**Snowflake's modern data lake solution**

According to [69], while Cloud storage providers like Amazon an Microsoft use Amazon Simple Storage Service (S3) and Microsoft Azure Blob to address the data swamps issue of Hadoop based data lakes, which enable users no more need to manage hardware stack and can increase virtual hardware easily, these users still need to manage their data, business and analytics applications, other essential activities on the clouds, and figure out how to maximize analytics performance.

Snowflake describes modern data lake as a place where data are stored with their raw form, no matter these are structured or semi-structured data. These object storage format of data enables modern data lake can store, load, integrate and analyse the stored data easily, and at the same time facilitate data driven decision making by deriving insights from the data.

Semi-structured and unstructured data may be the most valuable data, data lakes will ingest these kinds of data with a schema-as-read technology that makes near-real time analyse possible.

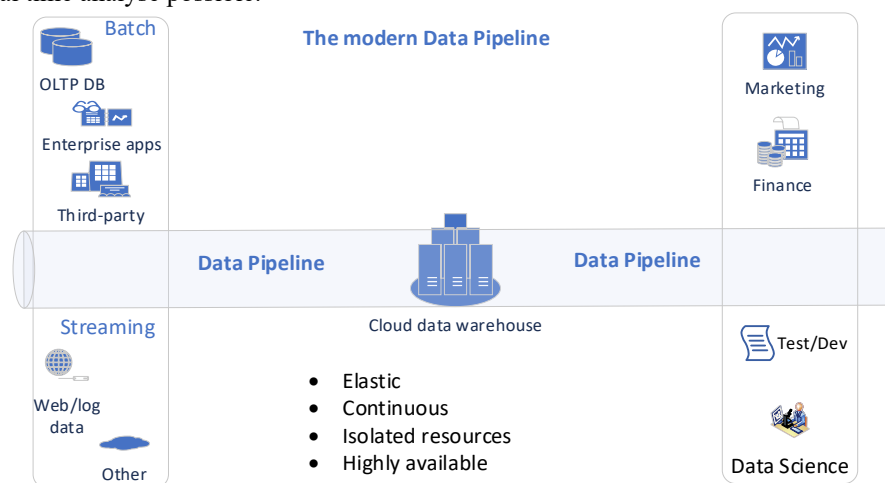

**Figure 2.8 Structure of modern data pipeline (reproduced from [69])**

The figure above shows a modern data pipeline structure. As can be seen from the figure, data from different sources in different formats can all be ingested into the data lake, in either batch or streaming manner. The data are all stored in the cloud data warehouse, and it can be stored in an external object store such as S3 or Azure Blog Storage for deeper and more effective analytics. This structure is elastic, can automatically and asynchronously detect and ingest new data in near real time manner, can isolate data resources and provide high available of data, as data are nearly immediately detected and ingested into the data warehouse for further analysis and mining.

**Cloud data warehouse**

Relational databases management systems (RDBMS) was the dominant technology for the data manage and analysis during 1980s when companies developed their own applications to solve data analysis issues, with the invention of relational data model according to Codd's principles [70]. As volume of data increasing and more users need access to the databases, competitions between users and data integration cause performance issues. Data warehouse appliance, which was a kind of black-box architecture that integrated hardware, software and networking components, emerged during 1990s to address data warehouse rigors. Then during 2000s, to address the rapid data growth issue, platforms and software such as Hadoop which was designed to support mixed and unpredictable workload of data warehouse were created. However, today's data sources are more numerous and varied, need support real-time or near real-time ingest in both steaming and batch processing manner, require increasing storage and computational capacity simply, timely and cost effectively. The new requirements motivated the generation of cloud data warehouse. The following figure (Figure 2.9) shows the evolution of the data warehouse from very beginning to the current version.

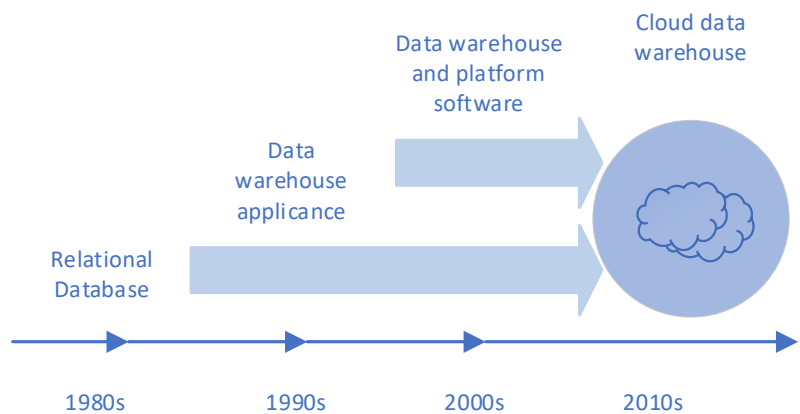

**Figure 2.9 Evolution of cloud data warehouse (Reproduced from [71])**

Technology features of cloud data warehouse include but not limited to the following:

**The could:** the essential technology enables the cloud data warehouse, makes the warehouse near-infinite, low-cost, high scalability, and pay for only the storage and usage.

**Massively parallel processing (MPP)**: split single computing task into many small tasks which can run simultaneously over separate computational units.

**Columnar storage**: data cell of a record is stored based on column, enables fast query of just one data element and thus improve query performance.

**Vectorized processing**: this feature is stemmed from the new hardware design which delivers fast data analytics performance compared with the tradition hardware technology

**Solid State Drivers (SSDs)**: this is another feature comes from the new hardware technology which stores data on flash memory chips that accelerates disk IO.

**Benefit of modern cloud data lake**

According to Snowflake [69], scalability, cost, productivity and simplicity are major benefits comparing with the traditional data lakes.

**Scalability**: on-premises data lakes require purchase and maintain hardware and software infrastructure, and need to plan hardware for peak processing loads, which may only happy annually. This results in either cost a lot on hardware, or struggling to handle peak processing loads. On contrast, modern cloud data lake providers separate data storage and computation services independently but locally integrated, and users just pay what they use. These cloud providers such as Amazon and Microsoft Azure can support virtually unlimited storage and computation capability, easy or automatic scale up and down to usage requirements without adversely degrade system performance.

**Deployment and Management Cost**

Tradition data lakes require also considering the deployment and management costs include system setup and tuning, security and daily management, data backup, and licence fees for software. On contrast, cloud vendors will automatically provide all the technique supports for all system management issues, and users thus can concentrate on extracting values from the data available on the cloud.

**Boosting Productivity**

Traditional data lakes need organizations to manage the system on-premise, so administrators must frequently involve capacity planning, resource allocation, performance tuning and other complex tasks that need highly skill experts to maintain the system running stably. On the other hand, modern cloud service providers usually take charge of all system management and maintain tasks, and consequently software engineers and analysts can spend more time on innovative projects.

**Simplifying the Environment**

Today, one of the challenges in big data is ingest data from heterogenous data sources in a timely manner into data storage infrastructures such as cloud. Data are generated continuously from either SQL or NoSQL databases, from Internet of Things (IoT) devices, SaaS and enterprise applications in structured, semi-structured, or unstructured formats. Cloud-based storage technology with object storage options can eliminate data solos and consolidate data in different formats into Azure Blob Storage or S3, without concerning software customization, programming, and system resources planning that are necessary in traditional data lakes. Users can then explore, analyse and visualise these consolidated data by using familiar tools such as SQL, or Python. As stated in [69], with the mature of pre-integrated cloud data lake technology, users can benefit from the relatively inexpensive object storage mechanisms, and use it immediately after sign up the services.

### 2.6 Enterprise Data Center/Cloud – Cloudera Data Platform

As the ever-increasing volume of data from heterogenous sources with variety formats, find insights from the data is a challenging issue for nearly all size of businesses, organizations, and enterprises. According to a report [72], 71 percent of a survey participants expressed completely or mostly agree with that they would consider to adopt an enterprise intelligent platform. Here, the Enterprise Intelligence Platform is defined as:

> *a single product or service that provides a superset of analytics/data science/data management functionality that is aimed not just at data consumers (e.g., data analysts, business analysts, data scientists) but also data operators (data management professionals and IT)*

The above definition implies three major components, which are historically three separate products from different vendors; 1) using ETL (Extract, Transform, and Load) products to ingest and integrate data from different sources and enterprises applications, 2) using traditional relational database management systems to store and processes structured data which is modelled and schema applied, and 3) using data science applications and business intelligence tools to analyse and visualise data.

Cloudera Data Platform is a typical such kind of Enterprise Intelligence Platform which we are going to introduce in the following section.

### 2.6.1 Overview of Cloudera Data Platform

According to Cloudera's official definition,

> *"Cloudera Data Platform (CDP) Data Center is the on-premises version of Cloudera Data Platform. This new product combines the best of Cloudera Enterprise Data Hub and Hortonworks Data Platform Enterprise along with new features and enhancements across the stack. This unified distribution is a scalable and customizable platform where you can securely run many types of workloads." [73]*

One key feature of CDP is that computation is separated from data storage. This is a kind of hybrid solution for the organizations and enterprises that require both cloud services and on-premises data centre. The hybrid approach can containerize applications[4] by managing storage, table schema, authentication, authorization and governance [73]. The following figure (Figure 2.10) demonstrates the main components of CDP. As can be seen from the figure, it supports three kinds of clouds, namely, multi public cloud, hybrid cloud, and data centre and private cloud.

---

[4] Containers are packages of software that includes everything that it needs to run, such as code, dependencies, libraries, and more. Container differs from Virtual Machines because container shares OS kernel rather than have a full copy of OS kernel for each VM.

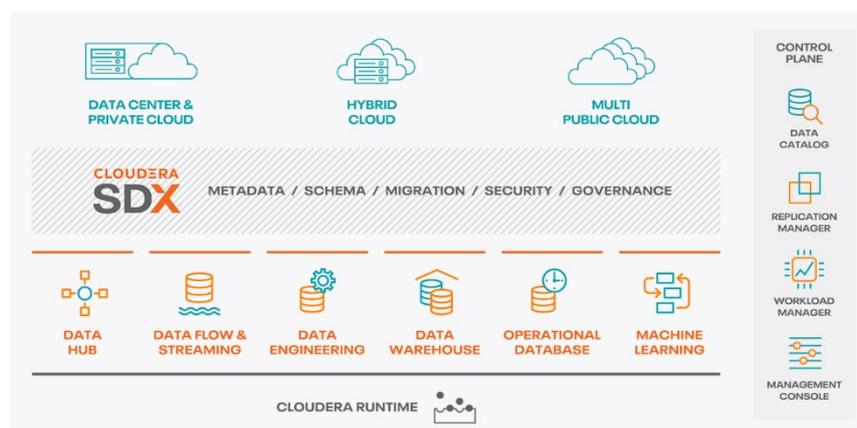

**Figure 2.10 Components of Cloudera Data Platform (source: https://www.cloudera.com/products/cloudera-data-platform.html)**

CDP public cloud service enables users to create and manage secure data lakes. The public clouds are managed by Cloudera and runs on AWS and Azure, but the data is under control by users. Other features include adaptive scaling which will match performance requirements of new workloads to speed up deployments; intelligent migration simplifies workloads migration for capacity requirements; burst to cloud facilitate organizations move data, metadata, or policies etc. to public cloud, and provide the right amount of capacity. The functions for public cloud are also provided for private cloud and hybrid cloud.

In the centre of CDP is Cloudera Shared Data Experience (SDX), which provides and deliver consistent data security and governance capabilities. SDX together with Control Plane, manage and deploy CDP services on the cloud. This integrated management framework responsible for the system security and governance, metadata, data catalog and data schema management, data migration, data replication and workload management. CDP has management console that allows system administers easily manage all clusters in CDP, including on-premises cluster, workloads, public clouds, private clouds, clusters in data centre, data warehouse, data hub and other CDP components, for which some of them will be discussed in the following sections.

CDP has a component called Cloudera Runtime that contains all necessary Cloudera components as listed below. Different components provide different services for different applications.

- Apache Atlas: a metadata store for metadata exchange both within and outside Hadoop stack.
- Data Analytics Studio: diagnostic tools for Apache Hive.
- Apache HBase: non-relational database for random and persistent access to tabular data.
- HDFS: Hadoop Distributed File System for data storage.

- Apache Hive: a data warehouse system includes Hive and Hive metastore for metadata storage.
- HWC: Hive Warehouse Connector, a tool to access and manage Hive table from Spark.
- Apache Hue: web-based query editor to interact with data warehouses.
- Apache Impala: enable high performance, low-latency SQL queries on HDFS data.
- Apache Kafka: a publish/subscribe like messaging system for streaming message processing
- Key Trustee Server: an enterprise-grade cryptographic key storage and management system.
- Key Trustee KMS: a custom Key Management Server (KMS) used by Hadoop KMS
- Apache Kudu: column-oriented data store for fast analytics on fast data.
- Apache Oozie: workflow scheduler and coordinate service for managing Hadoop jobs
- Apache Phoenix: a SQL layer for Apache HBase.
- Apache Ranger: a security component to control access to CDP with auditing and reporting service
- Apache Solr: a search engine for data stored in CDP
- Apache Spark: a distributed, in-memory data processing engine for large scale data processing
- Apache Sqoop: CLI (Command Line Interface)-based tool for bulk transfer of data between RDBMS and HDFS.
- Apache Tez: like Spark, Tez allows for a complex Direct Acyclic graph of tasks for processing data
- Apache YARN (Yet Another Resource Negotiator): the processing layer for managing distributed applications run on a cluster
- Apache Zookeeper: a distributed coordination service for managing large set of hosts

In the following sections, we will introduce Apache Hadoop HDFS which is the basic component for data storage; Apache Spark for fast in-memory parallel computing, Apache HBase for columnar data storage, and Apache Solr for unstructured data retrieval.

### 2.6.2 Hadoop HDFS

Let's first look at what is the definition of HDFS by Tome White (2015):

> *HDFS is a filesystem designed for storing very large files with streaming data access patterns, running on clusters of commodity hardware.[74]*

The definition itself reveals that HDFS is designed to address "*very large files*" up to petabytes in size. *"streaming data access"* implies HDFS is best suitable for the write-once, read-many-times processing pattern, although users can use Kudu to address fast-data (insert/update). In addition, "*commodity hardware*" means expensive, highly reliable hardware isn't necessary, instead, commonly available hardware on the market is a choice when build Hadoop Ecosystems. There is no Single Failure of Point (SPOF) issue in HDFS because one of the design objectives of HDFS is to address SPOF, so when a workstation is failure in a Hadoop cluster, there will be no noticeable interruption to users.

It's worth to note that HDFS itself may not a proper choice when an application requires *low-latency data access* (in the tens of milliseconds range), while HBase an Kudu are appropriate tools. If the data set contains *lots of small files*, say hundreds of millions small file, it will not a good choice to use HDFS to store the data because HDFS stores files to a block which is usually set to 128MB (by default) or 256MB, and consequently will cause waste of storage. Once again, HBase is a nearly perfect solution for small files and provide fast ad hock access by using row key. Last, if an application requires *multiple writers, arbitrary file modifications*, HDFS should not be used because HDFS assumes data are write once and read many times, so write is in an append-only fashion that does not a good solution for multiple writers' requirements.

Following are some basic concepts about Hadoop HDFS.

### Blocks

Different file systems have their own blocks which are minimum amount of data to read or write. Hard driver has its own block as well which is usually 512 bytes. The default HDFS block is 128MB or 256MB which is configurable. If the actually size of a file is less that the default block size, it will take the actual size on disk, but not the whole block. Using block, a very large file which cannot stored in a single disk can now be stored in a cluster with lot of disks. In addition, abstraction a block rather than a file makes the data storage easy to manage mainly because blocks are fixed size and easy to replicate on different data nodes.

### Data replication

HDFS stores large files across machines in a large cluster where each file is stored as a sequence of blocks, and the blocks are by default will be replicated three times in different computers in the cluster for fault tolerance. Number of replicas are configurable when a file is created and can be changed later.

### Namenodes and datanodes

An HDFS system operates on two types of nodes, namenode and datanodes. HDFS allows user data to be stored in files, then execute file operations such as opening, closing, and renaming files and directories. Namenode is similar to a master which stores metadata for all the files and directories in the HDFS filesystem tree. Users interact with HDFS via namenode by using a Linux-like Portable Operating System Interface (POSIX) which hides the storage details of the files.

Datanodes are workhouses of HDFS that store and retrieve blocks of files. The following figure illustrates the HDFS architecture with one namenode and two Racks in the cluster.

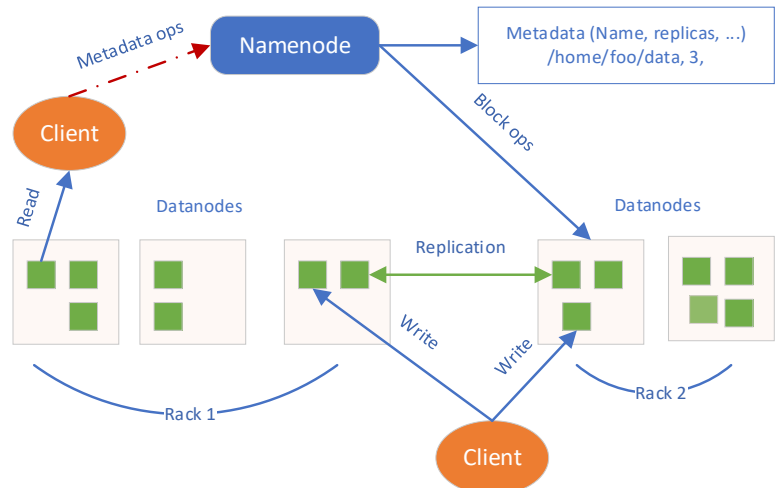

**Figure 2.11 HDFS architecture (reproduced from HHDFS Architecture [5])**

### HDFS Federation

If a cluster have to store a huge number of small files, and because namenode stores file metadata and block information in memory, memory may become a bottleneck for system performance. HDFS Federation allows to add more namenodes to a cluster, so different namenodes can manage files under different directories. Communication between namenodes are not necessary and failure of one namenode will not affect other namenodes.

### HDFS High Availability

HDFS uses secondary namenode to setup checkpoint to protect data loss, but when recovering data in the failure namenode, the HDFS service will stop during the restore procedure. To ensure high availability of HDFS, Hadoop allows to setup a pair of namenodes in an active-standby mode, and in case of active node is down, the standby node will become live node automatically without significant interruption. Upon recovering from failure, the previously down namenode will change to standby mode.

### Command Line Interface (CLI)

---

HDFS provides a list of POSIX-like commands to access files stored in HDFS. Common commands are listed below, refer to Hadoop File System Shell[6] for details.

- Make a directory
    hadoop fs -mkdir /user/hadoop/dir1, /user/hadoop/dir2
    hadoop fs -mkdir hdfs://nn1.example.com/user/hadoop/dir3
- List files in a given directory /user/hadoop/file1
    hadoop fs -ls /user/hadoop/file1
- List files in root directory
    hadoop fs -ls /
- List files in current user's directory
    hadoop fs -ls
- Copy file foobar.txt from local directory to current user's home directory in HDFS
    hadoop fs -put foobar.txt
- Copy file foobar.txt from current user's HDFS home to a given local directory
    hadoop fs -get foobar.txt /home/user1/localdir1
- Copy files from local to HDFS
    hadoop fs -copyFromLocal /home/user1/localdir/text1.txt /user/hadoop/dir1
- Download HDFS file to local
    hadoop fs -copyToLocal /user/hadoop/dir1/text1.txt /home/user2/localdir
- Print statistics about a file directory
    hadoop fs -stat "type:%F perm:%a %u:%g size:%b mtime:%y atime:%x name:%n" /file/user/hadoop/dir1
- Delete HDFS files
    hadoop fs -rm hdfs://nn.example.com/file /user/hadoop/emptydir
- Delete a HDFS directory
    hadoop fs -rmdir /user/hadoop/emptydir
- Get help information for an individual command
    hadoop fs -usage command

**HDFS Balancer**

It is not uncommon to add new datanodes to an existing Hadoop cluster. Then, namenode needs to consider a list of competing factors to place new blocks. The factors are[7]:

- Keep one of the replicas of a block on the same node as the node that is writing the block.
- Need to spread different replicas of a block across the racks.

---

[6]    https://hadoop.apache.org/docs/current/hadoop-project-dist/hadoop-common/FileSystemShell.html

[7]    https://hadoop.apache.org/docs/current/hadoop-project-dist/hadoop-hdfs/HdfsUserGuide.html

- One of the replicas is usually placed on the same rack as the node writing to the file.
- Spread HDFS data uniformly across the datanodes in the cluster.

To balance the storage on datanodes, HDFS provides a tool for system administers to rebalance the data across the cluster, and thus improve the performance of the Hadoop cluster.

### 2.6.3 Apache Hadoop YARN (Yet Another Resource Negotiator)

Apache YARN is a resource management system in a Hadoop cluster. It separates resource management from the job scheduling/monitoring into different daemons (processes or programs running in background). YARN contains a global Resource Manager (RM) and Application Maters (AM) for each application. The main components in YARN include:

- Resource Manager: runs on master daemon, manages resource allocation in a Hadoop cluster. Scheduler and Application Manager[8] are two components of RM. While scheduler purely responsible for allocating resources to various running applications; Application Manger accepts job-submissions, negotiating the first container for executing an AM, and restarting AM container on failure.
- Node Manager (NM): runs on each work nodes, executes and manages tasks on the nodes, monitors resources usage of individual container, sends heartbeats with health status with RM, and kills containers as directed by RM.
- Application Master: per-application AM requests resources from RM Scheduler for applications and manages the applications lifecycle. Work together with NM monitors the execution of tasks; it also sends heartbeats to RM to report health status.
- Container: a resource package incorporates RAM, CUP cores, network, Hard Driver etc. on a single node.

The following figure (**Figure 2.12**) illustrates the structure of YARN, and it also demonstrates how jobs are executed on YARN.

---

[8] To avoid confusion, we always use AM to represent Application Master, which is per application based, and use full name of Application Manager, which is component of Resource manager.

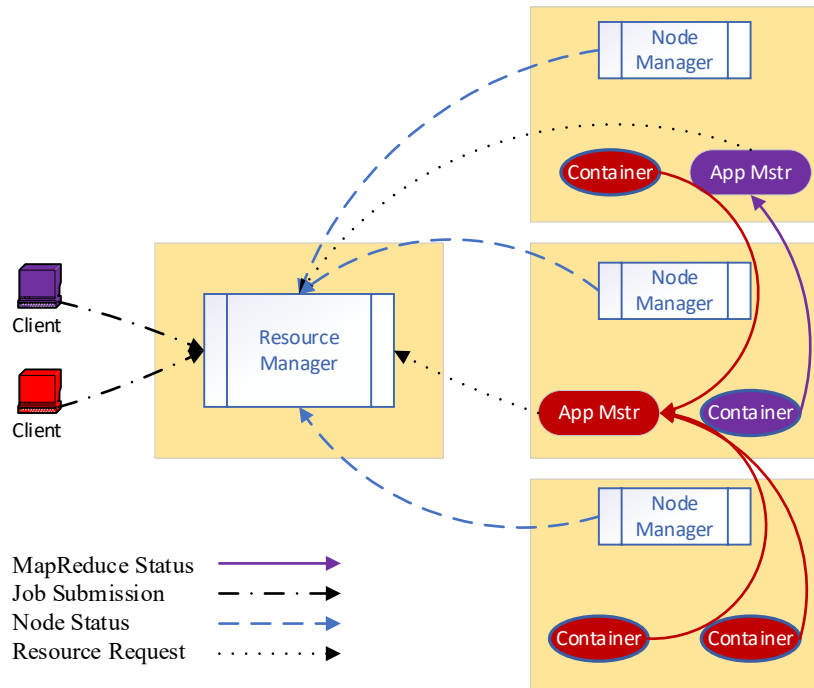

**Figure 2.12 Structure of YARN (reproduced from Apache Hadoop YARN[9])**

As shown in the figure, a Hadoop application is executed as following:

1) clients submit their jobs or applications to RM
2) Application Manager in RM allocates a container to start an AM (Application Master) for each application
3) AM then registers itself with RM on its status
4) When AM starts properly, it negotiates containers from Scheduler in RM for the application
5) When get the container list from RM, AM notifies NM to launche container for the application
6) The application will start to run on the containers, AM will track the run status and monitor the running progress
7) In case of failure of a container, AM will report the failure to RM, request a new container, then restart the job run on the failed container, the new container may in another datanode
8) In case of failure of AM, RM will kill the application, go back to step 2) to re-allocate a container for the killed job

---

9) Upon the success of the application, AM will return results to clients, and Application Manager of RM will responsible for unregistering the AM with RM.

## 2.7 Apache Spark

### 2.7.1 What is Apache Spark

Based on the official website of Apache Spark, Spark "is a unified analytics engine for large-scale data processing". Further, Spark provides high-level tools including Spark SQL for SQL and structured data processing, MLlib for machine learning, GraphX for graph processing, and Structured Streaming for incremental computation and streaming processing.[10] Tom White defined Apache Spark as "a cluster computing framework for large-scale data processing." [74] Although Spark does not use MapReduce as its execution engine, there are similarities in terms of API and runtime.

Some outstanding features of Apache Spark including but not limited to speed, ease of use, generality, and runs everywhere.

- Speed: According to the experiments from [75], Spark can 40X faster than Apache Hive for SQL queries, and 25X faster than MapReduced based machine learning programs. When running Logistic regression in Hadoop and Spark, Spark can even 100X faster than Hadoop because Spark employs a state-of-the-art DAG (Directed Acyclic Graph) scheduler, a query optimizer and a physical execution engine to optimize SQL queries.

- Ease of Use: Spark provides native bindings for Java, Scala, Python, R and SQL, and more than 80 high-level operators that facilitate to take some of the programming burdens off the shoulders of developers in a distributed environment to build parallel applications. It also allows users to utilize REPL (Read, Evaluate, Print and Loop) interactively from the Scala, Python, R, and SQL shells.

- Generality: Apache Spark consists of Spark SQL, Spark Streaming, MLlib machine learning package, and GraphX[11] for different applications with different processing types (batch, streaming, iterative, and interactive). These libraries can be integrated, combined seamlessly, and interoperate closely in the same application.

- Runs Everywhere: Apache Spark can run as a standalone platform, it can also, most often, run on YANR, Hadoop Apache Mesos[12], Kubernetes[13], or in the cloud, with the ability to access data stored in HDFS, Alluxio[14],

---

[10] https://spark.apache.org/docs/3.0.0-preview2/
[11] https://spark.apache.org/graphx/
[12] https://mesos.apache.org/
[13] https://kubernetes.io/
[14] https://www.alluxio.io/

Apache Cassandra[15], Apache HBase, Apache Hive and a long list of other data sources in different formats such as csv, parquet, json, and Avro, just to name a few.

### 2.7.2 Main Spark Components

The following figure shows the main components of Apache Spark.

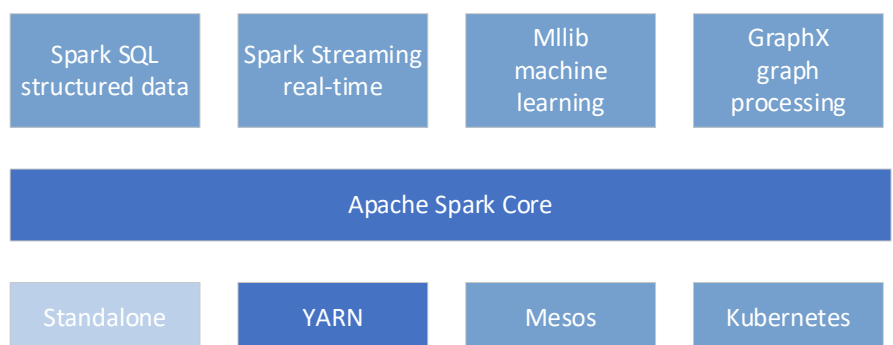

**Figure 2.13 The Spark Stack redraw from (reproduced from [76])**

#### Spark Core

As can be seen from the figure above, Spark core is a fundamental component that provides basic functions such as task scheduling, memory management, fault recovering, interacting with storage systems and other functions [76]. It also contains APIs that support Spark's programming abstraction – the Resilient Distributed Datasets (RDDs), which is an immutable collection of items distributed across servers in a cluster for fast and scalable parallel computing [76]. There are two types of operations on RDD, *transformations* and *actions*. First, an RDD can be created from simple text/csv/json/parquet/Avro files, SQL databases, NoSQL stores such as Cassandra or HBase, Amazon S3 or others. Once created, applying transformation will result in a new RDD from the previous one, for example, filtering data that matches a predicate. Actions in Spark will compute a result based on RDDs, either return it to the program calls it, or save the result to a file. Note that Spark only computes RDD transformations in a lazy fashion which means results are only computed when actions are called, and transformations of RDDs do not trigger immediate computations. This lazy computation fashion not only saves RAM but also takes minimum computations and thus can provide fast computation for big data [76].

#### Spark SQL

Spark SQL is a Spark module that integrates relational structured data processing with Spark's functional programming APIs. It allows SQL users and application programmers to intermix SQL queries with RDD transformations and actions and thus provide great flexibility with programming languages such as Java, Scala,

---

[15] https://cassandra.apache.org/

Python and R. According to [77], Spark SQL enables a tight integration between relational and procedural process; in addition, while keeping the Spark's programming model, the Catalyst, an extensible optimizer, greatly facilitates adding composable rules, controlling code generation and defining extension points for Spark software projects. Note that Spark SQL was originated from an open source project developed in University of California Berkeley, with the intention to provide a modified version of Apache Hive to allow it run on Spark.

### Spark MLlib

Spark MLlib is an open-source, distributed, scalable, and fast machine learning library which includes a framework for building machine learning pipelines. It fits into Spark APIs and can work together with Python on NumPy and R libraries. The library contains a variety of machine learning algorithms such as Logistic regression, Naïve Bayes, Decision trees, Random forest, k-means, Latent Dirichlet allocation, Association rules; and utilities a list of tools such as feature transformation, ML Pipeline construction, model evaluation and hyper-parameter tuning and others.

### Spark Streaming

The aim of Spark streaming is to provide a set of APIs that can easily handle and process live streams of data, such as web click information, log files generated by production web servers, and sensor data collected by IoT (Internet of Things) devices. Spark streaming APIs enable programmers to write streaming processing jobs similar to write batch jobs in Java, Scala and Python, while providing fault tolerance ability by recovering both lost work and operator state out of the box with no extra code needed. Spark stream allows not only batch processing, but also interactive ad-hoc queries on stream state. It supports read data from HDFS, Flume, Kafka, Twitter and a variety of other data sources. In production environment, it uses Zookeeper and HDFS for high availability [16].

### GraphX

GraphX provides the ability for Spark to manipulate graphs (e.g. social network relationship graph) and parallel graph computations. GraphX takes the advantages of parallel computation of Spark to achieve comparable performance to the special graph processing systems such as Apache Giraph[17] and GraphLab[18], while keeping the flexibility, fault tolerance, and highly flexibility APIs of Spark. Common graph algorithms such PageRank, Connected components, Label propagation, SVD++, Triangle count are all included in Spark's GraphX library.

### Cluster Managers

As stated earlier, spark can run on a cluster of servers to support parallel computing and scale out its computation capacity efficiently. This runs everywhere flexibility comes from the feature that Spark can run over not only its standalone

---

[16] https://spark.apache.org/streaming/
[17] https://giraph.apache.org/
[18] https://en.wikipedia.org/wiki/GraphLab

scheduler, but also platforms like Hadoop YARN, Apache Mesos, and Kubernetes containers.

### 2.8 Apache HBase

### 2.8.1 Concepts in HBase

HBase is a distributed column-oriented database built on top of HDFS, suitable for applications of large scale of stored data, and high I/O throughput random access to return a small subset of data [74]. Typical applications include Facebook messaging, webtable where key is a URL and value is metadata of the URL, geolocation like simple entities, graph data, sensor data and click counts. The characters of these applications include real-time random or batch access with small amount of query output, and the frequency of data update is high. However, keep in mind that if the application requires complex SQL queries, transactions, ACID compliance and multiple indexes on a table, HBase may not a good choice [74].

Before introducing HBase table, following are some concepts in HBase:
- Node: a single computer or server in a cluster
- Cluster: several nodes connected with each other and coordinated by a subset of the cluster
- Master Node: cluster coordinator
- Slave/Work node: execute jobs assigned by master node
- Daemon: a process or program runs in background
- Table: collection of rows
- Row: collection of column families
- Column: collection of key-value pair
- Namespace: Logical grouping of tables. HBase has two pre-defined namespaces, *hbase* for holding HBase internal tables; and *default* for tables without explicit namespace defined
- Cell: a {row, column, version} tuple exactly specifies a cell definition in HBase

A table consists of rows, columns (also known as qualifier) and column families which are created as part of the table definition. Each row has a unique row-key which likes a key in a map structure and is analogous to a primary key in a traditional RDBMS. Row-key is the main query field where rows are sorted and stored by. Data are hold in cells which are interactions of rows and columns. While this is similar to a tradition data table, but in HBase table, each column must belong to a column family; and all columns are organized by column families. HBase requires column family must be exist; but columns are created on the fly, and a column only exists when a given row has data in that column.

One significant difference between traditional RDMBS table with HBase table is that in HBase, a column family can have virtually any number of columns with the format of columnfamilyname:columnname. For example, an organization column family can have the columns *name* and *location* organization:name, organization:location, while an individual column family may have columns *firstname*, *lastname*, *sex*, and *nationality* which are expressed as individual:firstname, individual:lastname, individual:sex and individual:nationality. In practical, one can simply use a character to represent a column family to save storage, as shown below. In this table (Table 2.3), column family are presented in red, and column names are shown in blue.

**Table 2.3 Structure of HBase table**

| Row key | Org | | Ind | | | |
|---|---|---|---|---|---|---|
| | Nm | Loc | Fn | Ln | sx | nt |
| Cawafn1ln1fau | Companya | Wa | Fisrtname1 | Lastname1 | F | AUS |
| Casafn2ln2mus | Companya | Sa | Firstname2 | Lastname2 | M | USA |
| Cbnswfn3ln3men | companyb | Nsw | Firstname3 | Lastname3 | M | ENG |

HBase stores data on disk on a per-column family basis, and empty cells will not be stored to save space. In the above table, when stored in disk, data organized under column family *Org* will be stored in one file, and data belong to *Ind* column family will be stored in another separate file. In each of the separate files for different column families, the value of one cell is determined by Row key, column name in the format of columnFamily:columnName, and timestamp. For example, the value of USA (bold green in the last column in the above table) is determined by Casafn2ln2mus + Ind:nt + 1039493523294, here 1039493523294 is the timestamp when the cell is last updated.

Note also that column-family is the unit based on which HBase tunes and stores data, so it is advised members in a column family should have similar size and access pattern, for example, assign content of a picture in a separate column family, and assign metadata of the picture in another column family [74]. In addition, since HBase is designed for keeping short messages, the read and write operations are optimized for values smaller than 100KB, so the cell size should be kept relatively small - usually not be above 10MB and keep the total data size of a row no big than 100MB as a thumb of rule for performance purpose. In case of storing files bigger than 10MB, or 50MB when use mob[19], an alternative solution is to store the big file in HDFS and store the link to that file in HBase. The hard limit in a single table cell is 100MB and all values in a single row is 256MB[20].

### 2.8.2 Apache HBase Architecture

In a distributed environment, an HBase system contains several major components, or daemons include masters, region servers, Zookeeper quorum, HDFS NameNode and HDFS Datanodes. As shown in Figure 2.14, Master is a region server coordinator which in charge of bootstrapping a virgin install, creating new tables, assigning regions to registered region servers, monitoring region servers, recovering the failed region servers in case of failure, and other housekeeping operations. In a high available environment, there are multiple masters with one is active, and others as back masters.

---

[19] Mob: Medium-sized Objects,

[20] https://cloud.google.com/bigtable/docs/schema-design

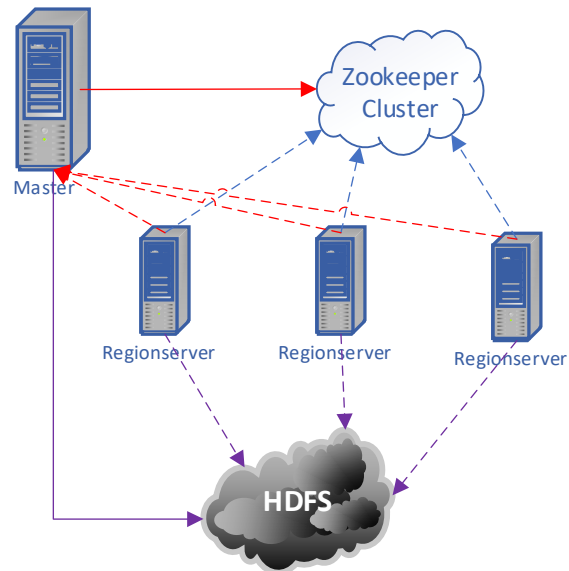

**Figure 2.14 HBase cluster member (reproduced from [74])**

Data stored in HBase tables are split into regions which are similar to sharding or partitioning in traditional RDBMs. Regions are denoted by the tables they belong to, and each region contains a subset of rows of a table. A region server services and manages several regions. After a table is created, it comprises a single region, and as the stored data growth, the original region is split into two new regions of approximately the same size, and this split process continues as the data growth, and the data are consequently stored in different servers in the HBase cluster. This mechanism allows one table can store data with hundreds or thousands or even millions of columns and billions of rows [74].

Zookeeper is a centralized component to coordinate services on all nodes, including both HBase masters and region servers, and maintain configuration information for HBase. HBase can have its own Zookeeper quorum or use the existing Zookeeper quorum. In case of multiple masters, the one which first connect to Zookeeper will become the active master and the rest are back or standby masters. If active master fails, one of the remaining masters will become controlling master, and the failed one, upon recover, will become standby master.

Namenode and Datanode are nodes in HDFS where the HBase data are stored.

### 2.8.3 Choosing a row key for a HBase table

Rows in HBase are sorted by row keys, and row key is the only indexed column in HBase, so it is extremely important to design row key to improve query performance and avoid hotspotting issue, which means huge client traffic is directed at one or small number of nodes in a cluster. The starting point of row key design to consider is how the data are accessed in business applications and put the most frequently queried data as part of row key, and at the same time, avoid sequential or time series row key design, because these designs may good for reading rows sequentially, but it seriously affects writing performance because all write will hit one region server. A row key with multiple fields may look like <sequential field><query field 1><query field 2>, here, sequential field is relatively small in length.

***Salted row key*** design moves a small, calculated hash at the beginning of a row key to randomize the row key, which can improve the write performance while still allows read by scanning which can ignore the salt field. Note that if row key is totally randomized by using MD5 or SHA256 hashing algorithms, it will greatly improve the write performance nevertheless read performance will be degraded dramatically because ad hoc query of the data must be rely on column field values which is the most inefficient operation in HBase [78].

Similar to salted row key strategy, ***promoted field key*** approach put a non-sequential or non-time series field in front of the row key. This strategy can also improve the write performance but may reduce the read/scan performance [78]. Therefore, there are balances between read and write performances in HBase, while sequence row key is preferable for read but not for write; random keys are more suitable for write but not a good choice for read intensive applications.

Given a concrete example, if one application is to find most recently data by country code, one can design row key as <country_code><Long.MAX_VALUE – System.currentTimeMillis()>, so the data in the same country will be stored in the same region, and most recent data will be sorted on the top of the table that will speedup retrieval of these most updated data.

### 2.8.4 Basic operations on HBase table from HBase shell

HBase shell is an interactive shell to allow user operate on HBase using Ruby syntax for commands, and pass parameters by using single quote (') and *hash rocket* notation such as in {PARMA => 'value'}. To start a HBase Shell, simply type in "hbase shell" in a Linux terminal. The following examples show some basic operations such as create a table, insert data into a table, query data, and drop a table.

- Create an HBase table mytable1 in namespace test, with two column family Org and Ind.

    *create 'test:mytable1', 'Org', 'Ind'*
- Insert rows into a HBase table

    *put 'mytable1', 'Cawafn1ln1fau', 'Org:Nm', 'Companya'*
    *put 'mytable1', 'Cawafn1ln1fau', 'Org:Loc', 'wa'*
    *put 'mytable1', 'Cawafn1ln1fau', 'Org:Typ', 'Gov'*

Notice that the last put command insert a new column "Gov" to the column family "Org" which is not shown in Table 2.3

- To retrieval a row by row key 'Cawafn1ln1fau' from table 'mytable1' in default namespace
    *get 'mytable1',* 'Cawafn1ln1fau'
- To get all rows from a table. Notice that if the table is huge, it may take a long time to finish
    *scan 'mytable1'*
- List all the tables
    *list*
- Get HBase status
    *status*
- Get number of rows in a table
    *count 'mytable1'*
- Get help information with simply typing in help, or *help command*
    *help*
- Drop a table, the table must first be disabled
    *disable 'mytable1'*
    *drop 'mytable1'*
- To delete all rows in a table but without delete the table itself, use truncate command
    *truncate 'mytable1'*
- Create a namespace
    *create_namespace 'test'*

Refer to Apache HBase Reference Guide[21] for more details of the HBase shell commands and other operations.

Note that in HBase shell, you can only use *put* to insert one cell value each time, not like by using HBase API, which you can use it to insert multi-cell values in one statement. Also note that there is no difference between insert and update in HBase, the put operation will always insert a new value to the cell with the row key, or update the value in that cell if the cell already exist.

### 2.8.5 Comparison of HBase with RDMBS

The following paragraph summarizes the features of HBase and RDBMS databases in general. It is worth to mention that HBase is not intended to replace existing RDMBS but it has its advantages in certain kinds of applications, which if served with traditional RDBMS, may suffer serious drawbacks as volume of data increase constantly.

---

[21] https://hbase.apache.org/apache_hbase_reference_guide.pdf

**Table 4 Features of RDBMS and HBase**

| Feature | RDBMS | HBase |
|---|---|---|
| Definition | Relational Database Management System. | Schema-less not-only relational database |
| SQL | Support SQL (Structured Query Language). | No support for SQL |
| Schema | Fixed, pre-defined schema | Column-family based dynamic column |
| Orientation | Row oriented | Column oriented |
| Scalability | Can scale up by using more powerful hardware | Highly linear scalable by add more servers in the cluster |
| Nature | Static in nature | Dynamic in nature |
| Data retrieval | An ID based retrieval involves many joins may slow | Row-key based retrieval is most effective |
| Rule | Support ACID | Follows CAP[22] (Consistency, Availability and Partition-tolerance) |
| Table normalization | Tables are normalized | Un-normalized flatten table, use de-normalization to avoid join operation |
| Transaction | RDMBS are transactional | No transactions exist in HBase |
| Data structure | Most suitable for structured data | Can handle unstructured, semi-structured and structured data |
| Spare data handling | No special handling | Can handle spare data well |

### 2.9 Apache Solr

### 2.9.1 What is Solr

Apache Solr is an open source enterprise search service build on top of Apache Lucene. It is a popular, blazing-fast, highly reliable, scalable and fault tolerant search service, providing distributed indexing, replication and load-balanced querying, automated failover and recovery and more. In 1999, Doug Cutting, the author of Apache Hadoop and Apache Nutch[23], wrote the Lucene as an indexing and searching engine for Nutch. In 2004, Solr was developed at CNET[24] by Yonik Seeley, and denoted to Apache Software Foundation in 2006. Since then, Solr has been developed by an active community, and has been used by AT&T, eBay, Instagram, Netflix, Disney, eHarmony, DuckDuckGo, Zappos.com, BestBuy and others. Now, Lucene and Solr are merged as a single Apache project[25].

---

[22] https://en.wikipedia.org/wiki/CAP_theorem

Solr is also the default search engine used by Cloudera for its data indexing and searching service.

### 2.9.2 Features of Solr

Following are a list of features of Solr. Keep in mind that Solr is a standalone enterprise search service which indexes different types of documents and then allows users to retrieve the indexed documents by using search terms just like when searching the Internet using Google. Some features of Solr are listed below:

- Advanced full-text search: build on Lucene, Solr provides different type of matching such as phrases, wildcards, joins, and grouping.
- Near real-time indexing: Solr can index content by using Lucene's real-time indexing capabilities and then make the content searchable.
- Extensible plugin architecture: with well-defined extension points, developers and users can easily add their own index and query plugins.
- Optimized for high volume traffic: Solr has been proven at extremely large scales the world over.
- Highly scalable and fault tolerant: benefited by combining zookeeper to facilitate highly scalable, fault tolerant distributed infrastructure.
- Comprehensive administration interfaces: Solr's built-in admin GUI allows users easily to control Solr instances.
- Geospatial search: Solr supports location-based search.
- Facet search and filtering: indexed data can be sliced and diced by facet searching.
- Query suggestion and spelling checking: Solr support Google like query suggestions and spelling checking.
- Rich document parsing: using Apache Tika[26], Solr can index rich context types such as Microsoft office documents, PDF files, XML, CSV, JSON, emails, and whatever Tika can parse.
- Security: Solr employs SSL, authentication and Role based authorization to secure the data indexed by Solr.
- SolrCloud: a cluster of Solr servers that provides distributed indexing and searching together with Apache Zookeeper.

---

[23] http://nutch.apache.org/
[24] https://en.wikipedia.org/wiki/CNET
[25] https://lucene.apache.org/
[26] https://tika.apache.org/

### 2.9.3 How Solr works

The following figure, which is reproduced from Solr Ref Guide 8.5[27], illustrates how SolrCloud integrated with a user application.

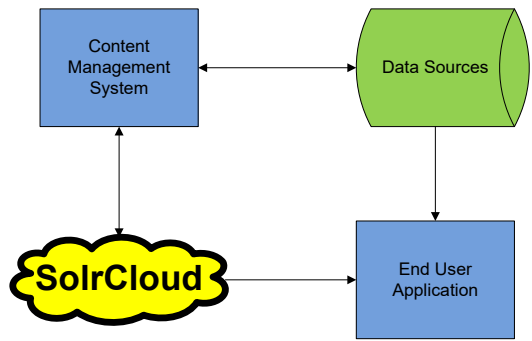

**Figure 15 Solr integration with applications**

There are four components in the above figure. SolrCould is running as a standalone indexing and searching service. An end user application, for example, an online store application, allows users shopping products from the online store. Content Management System provides an internal interface for store employees to update products information which may be obtained from different data sources. The metadata of the products will be indexed by Solr for end user consumption via simple HTTP request, and JSON, or XML or CSV as response format.

There are several steps to take before an application can access the data indexed by Solr, the section below will present a tutorial based on the steps to show how to create a SolrCloud search engine, and how to query the indexed data using Solrj, a Solr's Java API library.

### 2.9.3 SolrCloud setup – a tutorial

Solr expands its indexing and searching capability by using SolrCloud, which provides a highly scalable, available, and fault tolerant environment powered by Apache Zookeeper which provides coordination service of the working nodes in the SolrCloud cluster. This section will describe the detailed steps on how to setup a SolrCloud search engine.

We are using Ubuntu 18.04 as our operating system. Before starting the tutorial, make sure Java 8 or above is installed on Ubuntu.

**Step 1. Download and deploy Solr as a standalone service**

From https://lucene.apache.org/solr/downloads.html, download the newest binary release of Solr. It is Solr 8.5.0 when the chapter is written. If download source

---

release, one needs to compile Solr with Apache ant[28]. Note, you are not required to do anything from Step 2 to Step 4 for this tutorial.

~$ ls solr*
~$ solr-8.5.0-tgz
~$ tar zxf solr-8.5.0-tgz.gz
~$ cd solr-8.5.0/

**Step 2. Design schema for the data**

We are going to use a dataset distributed with Solr release. The collection of documents in located at ../example/exampledocs, which include 20 files in different formats such as JSON, CSV.

These kinds of files are called semi-structured data. To index the documents with predefined fields, we need to inform Solr about the characteristics of the fields in its configuration file, managed-schema.xml which is located at ../server/solr/configsets/sample_techproduts_configs/conf. Solr already defined all the related fields for these documents so there is no need to make any changes for it. In your own applications, change the fields to make them match the fields for the documents to be indexed and searched.

**Step 3. Configure parameters in solrconfig.xml**

Similar to change parameters in schema.xml, parameters in solrconfig.xml (located in the ../conf directory) allow users to control important features and behaviors of Solr. These include Request Handlers which process the requests to Solr; Listeners which listen for particular query-related events; Request Dispatcher for managing HTTP communications, and others.

Once again, Solr provides default settings for all the parameters when the sample data are indexed so for this tutorial, there is no need to change any of the parameters, but in your own application, you can change the settings to satisfy your own application requirements

**Step 4. Upload configurations to Zookeeper**

In a production environment, Zookeeper may have already run on different servers. To use the existing Zookeeper service, all the configuration files need to be uploaded to Zookeeper before starting SolrCloud service. In addition, to avoid confusion between Solr related configuration files stored in Zookeeper with other existing configurations for other applications, it is better to create an application specific path for different applications. For SolrCloud, we can create a path /solr in Zookeeper, and upload Solr configurations to that folder.

Solr provides a list of Zookeeper related commands to facilitate all the operations with regards to Zookeeper. The following command will create a znode by specifying a path for Solr by using Solr's Zookeeper command. In the command, we assume that Zookeeper is running on three servers, namely, server1, server2 and server3 using port 2181.

---

[28] https://ant.apache.org/

~$ bin/solr zk mkroot /solr -z server1:2181,server2:2181,server3:2181

Then, we can upload configuration files located at /path/to/solr/config to Zookeeper under the folder /solr, which we just created. Note, myNewConf is the name for this configuration.

~$ bin/solr zk upconfig -z server1:2181,server2:2181,server3:2181 -n myN-ewConf -d /path/to/solr/config

**Step 5. Starting SolrCloud services**

Now, with the configuration files have been uploaded to Zookeeper, it's time to launch Solr service in SolrCloud mode. To simplify the setup, we will use Solr's built-in Zookeeper ensemble, and use all the pre-defined configurations such as schema.xml and solrconfig.xml, so after finishing Step 1, readers can jump directly to this step.

The following command will 1) starts SolrCloud service on your local workstation, 2) asks how many Solr node will run on your computer, and 3) asks the ports of Solr (default 8983). At this point, SolrCloud service should start successfully.

~$ ./bin/solr start -e cloud

When prompt the following question, simply hit enter to use default values.

"… how many Solr nodes would you like to run in your local cluster? (specify 1-4 nodes) [2]:"

"Please enter the port for node1 [8983]:"

"Please enter the port for node2 [7574]:"

Then, Solr will shows the following information to indicate cluster is ready.
"Cluster at localhost:9983 ready"

**Step 6. Create a collection**

Before the data can be indexed, Solr needs to create a collection for the dataset. Same as before, since we are using SolrCloud, we can use bin/solr scripts to create the collection. The following command will create a collection myCollection1, with the configuration named as myNewConf created at Step 4, Solr port is 8983, create the collection on 3 shards, with replica 2.

~$ bin/solr create -c myCollection1 -n myNewConf -p 8983 -s 3 -rf 2

NOTE: You don't need to execute the above command unless you are using your existing Hadoop cluster with more than servers

Continue to Step 5, after Solr shows "Cluster at localhost:9983 ready", it will try to automatically create a new collection for you by prompt

"Please provide a name for your new collection: [gettingstarted]"

Hit Enter key to accept the default collection name. Then Solr will ask you number of shards of the collection and number of replicas

"How many shards would you like to split techproducts into? [2]"
"How many replicas per shard would you like to create? [2]"

Hit Enter again to accept the default values for the above two questions.

The last question Solr will ask is as following:
"Please choose a configuration for the techproducts collection, available options are:
 _default or sample_techproducts_configs [_default]"

Since we are going to use sample_techproducts_configs for the tech products dataset, so type in sample_techproducts_configs for this question.

Solr now has all the necessary information, that is, name of a collection, number of shards and number of replicas for the collection, to start SolrCloud service. It will output the following information to show the status of Solr:
"SolrCloud example running, please visit: http://localhost:8983/solr"

In a browser, visit http://localhost:8983/solr to access SolrCloud admin interface. Click the Cloud tab on the left panel will show a diagram like following.

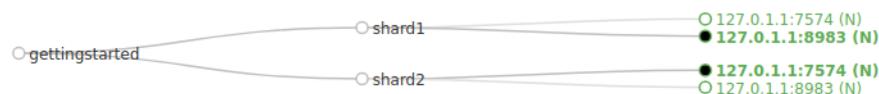

**Figure 16 SolrCloud diagram**

### Step 7. Index data
Now, everything is ready for SolrCloud to index data. Solr includes a script that facilitate the indexing procedure. Run the following command will make SolrCloud indexed the data stored at ../example/exampledocs/* to collection gettingstarted.
~$ bin/post -c gettingstarted example/exampledocs/*

**Step 8. Query data**

Now data are indexed, and users can search the data via REST clients, curl, wget, Chrome POSTMAN, etc, or via programming languages such as Java, Python, Scala, PHP, and others. Solr admin user interface also allows users to search the indexed data for debugging purpose.

The following code shows how to use Solrj, a Java-based client side API for Solr, to search the indexed data in the previous steps. It uses "foundation" as search term, and prints out how many results returned by SolrCloud.

For a maven project, need to include the following in your pom.xml file

```
<dependency>
    <groupId>org.apache.solr</groupId>
    <artifactId>solr-solrj</artifactId>
    <version>8.5.0</version>
</dependency>

String url = http://localhost:8983/solr/techproducts;
HttpSolrClient solr = new HttpSolrClient.Builder(url).build();
Solr.setParser(new XMLResponseParser());

SolrQuery query = new SolrQuery();
query.set("q", "foundation");
QueryResponse response = solr.query(query)
SolrDocumentList results = response.getResults();
System.out.print("# of results: " + result.getNumFound());
```

Solrj can perform nearly all of the operations discussed above that allows a Java application to interact with Solr with simple high-level methods.

## 2.10 Resources

Following are some resources for practitioners and researchers on big data.

Top 10 references for applying big data and analytics in business: https://www.kdnuggets.com/2014/08/top-10-references-applying-big-data-analytics-business.html

Wikepedia: https://en.wikipedia.org/wiki/Big_data

Big Data Research
https://www.journals.elsevier.com/big-data-research

Big data analytics references

https://datascience.stackexchange.com/questions/15260/big-data-analytics-references

Books
- Big Data and Analytics (WIND)
- Hadoop for Dummies
- Big Data for Dummies
- Hadoop: The Definitive Guide
- Learning Spark: Lightning-Fast Big Data Analysis
- MapReduce Design Patterns

Online Courses
- Udemy - Big Data Analytics Courses
- Coursera - Big Data Analytics Courses
- edX - Big Data Analytics Courses
- Udacity - Data Analysis

Update [BigData in R]
- Wikipedia - Programming with Big Data in R
- RStudio - Working with BigData in R
- InfoWorld - Learn To crunch BigData with R

Big Data And AI: 30 Amazing (And Free) Public Data Sources For 2018
https://www.forbes.com/sites/bernardmarr/2018/02/26/big-data-and-ai-30-amazing-and-free-public-data-sources-for-2018/#4321299d5f8a

Big Data Analysis Learning Resources: 50 Courses, Blogs, Tutorials, and More for Mastering Big Data Analytics
https://www.ngdata.com/big-data-analysis-resources/

Big Data Resources
https://www.dummies.com/programming/big-data/big-data-resources/

9 Online Resources to look for Information on Big Data
https://codecondo.com/9-online-resources-to-look-for-information-on-big-data/

## 2.11 Conclusion

This chapter provides an overview to the Big data technology that is commonly utilised to construct social big data applications. In particular various currently incorporated technologies, tools, APIs, and approaches are discussed that are used from infrastructure/platform/ecosystem to constructional units and components. Also, an array of online resources is provided to help researchers and professionals to obtain further and deep insights into Big data technology.